\documentclass[preprint,aps]{revtex4}

\usepackage{graphicx}
\usepackage{dcolumn}
\usepackage{bm}

\begin{document}


\title{On the influence of hidden momentum and hidden energy in the classical analysis of spin-orbit coupling in hydrogenlike atoms}

\author{David C. Lush } 
\affiliation{%
 d.lush@comcast.net \\}%


\date{\today}



\begin{abstract} 

In a recent article, Kholmetskii, Missevitch and Yarmin [``On the classical analysis of spin-orbit coupling in hydrogenlike atoms,'' Am. J. Phys. 78(4), April 2010 (428-432)] examine in detail the spin-orbit coupling in the semiclassical hydrogenic atom, and identify a need to account for non-Coulomb forces not included in the standard analysis.  Kholmetskii, {\em et al.}, showed that the experimentally-measured coupling continues to be obtained when the new forces are incorporated in the analysis. This requires that the change in orbital radius due to non-Coulomb forces is also properly accounted for.  In response to a comment, Kholmetskii, {\em et al.}, showed that the experimentally-measured coupling continues to be obtained when so-called ``hidden momentum'' forces are also included.  However it has been postulated that when hidden momentum is nonvanishing, a corresponding hidden energy must also be present, that was not included in the total energy by Kholmetskii, {\em et al.} Hidden energy is postulated necessary to obtaining a relativistically covariant description. Inclusion of hidden energy leads to disagreement with the experimentally-determined spin-orbit coupling magnitude if the Bohr postulate, that orbital angular momentum is quantized in whole multiples of the reduced Planck constant, is assumed to apply to kinetic momentum alone.  The empirical result may be recovered in the semiclassical picture and using the general approach developed by Kholmetskii, {\em et al.}, if the Bohr postulate is reinterpreted to apply to orbital angular momentum that consists of hidden as well as kinetic momentum.

\end{abstract}

\maketitle

\section{Introduction}

In a recent article, Kholmetskii, Missevitch and Yarmin \cite{KholmetskiiY10} examine in detail the spin-orbit coupling in the semiclassical hydrogenic atom, and identify a need to account for certain forces not included in the standard analysis, due to translational motion of the magnetic electron.  In a comment on the article \cite{LushY10}, it was observed that an additional force was unaccounted for, due to the existence of time-varying ``hidden'' momentum of the electron intrinsic magnetic dipole in the electric field of the nucleus.  Hidden momentum refers to mechanical momentum of a current carrying body in an electric field, that is nonvanishing in the rest frame of the body, and is separate from the kinetic momentum due to the motion of the body.  (For more detailed description of hidden momentum, please refer to \cite{ShockleyJames1967,Coleman1968,Hnizdo:1992}. For discussion of the necessity of accounting for hidden momentum in the semiclassical analysis of the hydrogen atom spin-orbit coupling, see \cite{MunozY1}.) 

In their reply to the Reference 2 comment, Kholmetskii, {\em et al.} \cite{KholmetskiiY10a}, showed that the experimentally-measured spin-orbit coupling magnitude continues to be obtained by the method of their paper, when modified to include hidden momentum forces.  However, it has been postulated \cite{Hnizdo:1997} that when a body possessing hidden momentum is in motion, a corresponding energy (sometimes termed ``hidden'' energy \cite{MunozY1oc}) must also be present in a relativistically-covariant electrodynamic description. It will be shown herein, inclusion of hidden energy leads to disagreement with the experimentally-determined spin-orbit coupling magnitude in the semiclassical picture, if the orbital angular momentum that is quantized omits the hidden orbital angular momentum. The empirical result is recovered in the semiclassical picture and using the method of Kholmetskii, {\em et al.}, if the Bohr postulate that orbital angular momentum is quantized in whole multiples of the reduced Planck constant, is assumed to apply to orbital angular momentum that consists of hidden as well as kinetic momentum.

\section{Hidden energy of the spin-orbit coupling compared to the measured coupling}

In this section it is shown that the hidden energy term in the hydrogenic energy calculation is of the same order as the spin-orbit coupling, and so cannot be neglected in an accounting of the total system energy.  For simplicity, consideration will be restricted to the case of the semiclassical Hydrogen atom in the Bohr model ground state, {\em i.e.}, for \(n=Z=1\) in the notation of Kholmetskii, {\em et al.} in Reference 1. 

The hidden momentum of a magnetic dipole \mbox{\boldmath$\mu$}  in an electric field \mbox{\boldmath$E$} is,  in Gaussian units here, \(\mbox{\boldmath$P$}_{\text{hidden}} = (\mbox{\boldmath$\mu$} \times \mbox{\boldmath$E$})/c\) \cite{Hnizdo:1992oc}.  The hidden energy of a magnetic dipole translating with velocity \(\mbox{\boldmath$v$}\) is given \cite{Hnizdo:1997oc} generally as

\begin{equation}
H =  \gamma \mbox{\boldmath$v$}\cdot\mbox{\boldmath$P$}_{\text{hidden}} = \gamma \mbox{\boldmath$v$}\cdot \left[\frac{\mbox{\boldmath$\mu$}\times\mbox{\boldmath$E$}}{c}\right]
\label{HiddenEnergyGeneral}
\end{equation}

where \(\gamma \equiv (1-(v/c)^2)^{-1/2}\) with \(v \equiv |\mbox{\boldmath$v$}|\) and \(c\) the speed of light.  For an electron of mass \(m\) and charge magnitude \(e\) in a circular orbit of radius \(r\) around an arbitrarily heavy proton, under the influence of only the Coulomb force, \(v = e/\sqrt{mr}\) \cite{HallidayResnick}.  Since the non-Coulomb force due to the translational motion of the magnetic electron is small compared to the Coulomb attractive force at the atomic scale of interest, the resulting change in the electron orbital velocity compared to its motion under Coulomb attraction alone is expected to be small. The change in the electron velocity due to the non-Coulomb force can therefore be neglected in establishing the significance of the hidden energy (although it will be evaluated explicitly in what follows).     

For simplicity and following Kholmetskii, {\em et al.}, consideration here is restricted to cases where the electron intrinsic magnetic moment is perpendicular to the orbital plane.  Since the electric field due to the proton (\(\mbox{\boldmath$E$} = e \mbox{\boldmath$r$} / r^3\) here, where \(\mbox{\boldmath$r$}\) is the radius vector from the proton to the electron) is outwardly radial, and since the electron magnetic moment is oppositely directed to its intrinsic spin angular momentum vector, it can be seen from Eq. (\ref{HiddenEnergyGeneral}) that a negative value of \(H\) results for co-directional spin and orbital angular momenta.  With the electron intrinsic magnetic moment magnitude assumed equal to the Bohr magneton \(e\hbar/2 m c\), and approximating the velocity as \(v = e/\sqrt{mr}\) and \(\gamma\) as unity, the hidden energy of Eq. (\ref{HiddenEnergyGeneral}) becomes

\begin{equation}
H_{\pm} \approx \mp \frac{e}{m^{1/2}r^{1/2}}\frac{e \hbar}{2 m c^2} \frac{e}{r^2}= \mp \frac{e^3 \hbar}{2 m^{3/2} c^2 r^{5/2}},
\label{V_a}
\end{equation}

where \(\hbar\) is the reduced Planck constant.  Here, and henceforth herein, the upper (lower) sign represents parallel (antiparallel) spin and orbital angular momenta.  At the Bohr radius \( r_{\text{B}} = \hbar^2/e^2m\) this becomes

\begin{equation}
H_{\pm} \approx \mp \frac{e^8 m}{2 c^2 \hbar^4}.
\label{H_pm}
\end{equation}

To see that the hidden energy is of similar magnitude to the spin-orbit coupling as measured experimentally, it can be recalled that the measured value can be obtained (see for example Eq. (15) of Reference 1 and the preceeding discussion) as 

\begin{equation}
V_{s-o} = -\frac{\mbox{\boldmath$\mu$}\cdot\mbox{\boldmath$B$}}{2}
\label{V_so_ef1}
\end{equation}

where \(\mbox{\boldmath$B$} \) is the magnetic field at the electron in the electron rest frame. Noting that the proton velocity in the electron rest frame is \(-\mbox{\boldmath$v$}\), where \(\mbox{\boldmath$v$}\) is the electron velocity in the laboratory frame, the magnetic field in the electron rest frame is \(\mbox{\boldmath$B$} = (e(-\mbox{\boldmath$v$}) \times \mbox{\boldmath$r$})/c r^3 =  (e\mbox{\boldmath$r$} \times \mbox{\boldmath$v$})/c r^3 =  e\mbox{\boldmath$L$}/m c r^3\).  Thus

\begin{equation}
V_{s-o} = -\frac{\mbox{\boldmath$\mu$}\cdot\mbox{e\boldmath$L$}}{2 m c r^3}, 
\label{V__so_ef1}
\end{equation}

where \(\mbox{\boldmath$L$} \equiv\mbox{\boldmath$r$} \times \mbox{\boldmath$P$} = \mbox{\boldmath$r$} \times m \mbox{\boldmath$v$} \) is the kinetic orbital angular momentum.  (The effect of the Thomas precession in reducing the proton velocity magnitude has been neglected here, but this is only an order \((v/c)^2\) error in an order \((v/c)^2\) correction to the energy, and so is negligible to the present ananlysis.) At the Bohr radius where \(L \equiv |\mbox{\boldmath$L$}| = \hbar \),

\begin{equation}
V_{s-o \pm} = \pm \frac{\mu e \hbar}{2mc{r_{\text{B}}}^3}.   
\label{V__so_ef2}
\end{equation}

Replacing \(\mu\) by its value as the Bohr magneton and the Bohr radius by its formula obtains

\begin{equation}
V_{s-o \pm} = \pm \frac{e\hbar}{2mc}\frac{e \hbar}{2 m c} \left(\frac{me^2} {\hbar^2}\right)^{3}  = \pm \frac{e^8 m}{4c^2\hbar^4}.
\label{V_so_final}
\end{equation}

Comparison of Eq. (\ref{V_so_final}) with Eq. (\ref{H_pm}) reveals that the hidden energy contribution to the total system energy is twice the magnitude and of opposite sign to the measured coupling.  Therefore hidden energy is not negligible in an assessment of the spin-orbit coupling.  Since Kholmetskii, {\em et al.} show that the spin-orbit coupling taking account hidden momentum but omitting hidden energy is in agreement with experiment, accounting for hidden energy as well as hidden momentum is seen to change the sign of the spin-orbit coupling for each electron spin orientation.  This would not be readily spectroscopically observable in the absence of an applied field, but would lead to  disagreement with the observed Zeeman effect under an externally applied magnetic field.  

\subsection{General relationship between orientational and hidden energy}

For the electric field at the electron due to the proton charge, Equation (\ref{HiddenEnergyGeneral}) can be rewritten generally as

\begin{equation}
H = \frac{\gamma e}{cr^3} \mbox{\boldmath$v$}\cdot \left(\mbox{\boldmath$\mu$}\times\mbox{\boldmath$r$}\right) = \frac{\gamma e}{cr^3} \mbox{\boldmath$\mu$}\cdot \left(\mbox{\boldmath$r$}\times\mbox{\boldmath$v$}\right).
\label{HiddenEnergyGeneral_extended}
\end{equation}

The potential energy of the relative spin-orbit orientation of Eq. (\ref{V__so_ef1}) can be rewritten similarly as

\begin{equation}
V_{s-o} = -\frac{\mbox{\boldmath$\mu$}\cdot\mbox{e\boldmath$L$}}{2 m c r^3} = -\frac{e}{2 c r^3}\mbox{\boldmath$\mu$}\cdot(\mbox{\boldmath$r$} \times \mbox{\boldmath$v$}).
\label{V__so_ef1_copy}
\end{equation}

So, apart from the \(\gamma\) factor, which differs from unity maximally at the ground state radius and there only by about \(10^{-5}\),

\begin{equation}
V_{s-o} = -H/2.
\label{H_to_vso_general}
\end{equation}

The relationship found above, specifically for the ground state Bohr radius and spin orientations perpendicular to the orbital plane, thus holds for general Bohr level and relative orientation of the spin and orbit.

\section{Incorporation of hidden momentum in the Bohr second postulate}

The Bohr second postulate, that the orbital angular momentum magnitude \(L\) is quantized in whole multiples \(n\) of the reduced Planck's constant, is expressible as

\begin{equation}
L = rmv = n \hbar. 
\label{BohrPostulate}
\end{equation}

It is worth noting that in order to obtain the best agreement between the Bohr model and experiment, the mass \(m\) here must be the electron reduced mass \(m_r \equiv m_e m_p /(m_e + m_p)\), where \(m_e\) and \(m_p\) are the electron and proton masses. When the Bohr postulate is so expressed in terms of the reduced mass, it is postulating that it is the total orbital angular momentum of the electron and proton that is quantized, rather than just the electron orbital angular momentum alone.  Therefore it should not be considered completely without basis to suppose that hidden orbital angular momentum should be explicitly included in the Bohr second postulate.  Rather, it can be taken as a straightforward consequence of supposing that the total orbital angular momentum is the quantity quantized in whole multiples of \(\hbar\). 

Incorporating hidden orbital angular momentum, the Bohr second postulate can be expressed as

\begin{equation}
L_T \equiv |\mbox{\boldmath$L$} + \mbox{\boldmath$L$}_{\text{hidden}} | = n \hbar,
\label{ModifiedBohrPostulate}
\end{equation}

where \(\mbox{\boldmath$L$}_{\text{hidden}} \equiv \mbox{\boldmath$r$} \times \mbox{\boldmath$P$}_{\text{hidden}}\) is the hidden orbital angular momentum.  For the circular orbit of arbitrary radius and assuming the electron spin is oriented perpendicularly to the orbital plane, the hidden orbital angular momentum magnitude is 

\begin{equation}
L_{\text{hidden}} \equiv |\mbox{\boldmath$L$}_{\text{hidden}}| = \frac{r E \mu}{c} = \frac{e^2 \hbar}{2 m c^2 r}. 
\label{L_hidden}
\end{equation}

Based on the definitions of the hidden momentum and hidden orbital angular momentum, if the spin angular momentum is co-directional with the kinetic orbital angular momentum, then the hidden orbital angular momentum is directed oppositely to the kinetic orbital angular momentum.  The modified Bohr second postulate is then expressible as

\begin{equation}
L + L_{\text{hidden}\pm} = r m v \mp \frac{e^2 \hbar}{2 m c^2 r} = n \hbar,
\label{ModifiedBohrPost}
\end{equation}

with \(L = rmv\) and \(L_{\text{hidden}\pm}\equiv \mp L_{\text{hidden}}\). This can be expressed in a more convenient form, with restriction to \(n=1\), as 

\begin{equation}
r m v = \hbar(1 - \delta_{\pm}), 
\label{ModBohrPostSimplified}
\end{equation}

with 

\begin{equation}
\delta_{\pm} = \delta_{\pm}(r) \equiv  \frac{L_{\text{hidden}\pm}}{\hbar} = \mp \frac{e^2}{2 m c^2 r}.
\label{delta_pm_def}
\end{equation}

For Hydrogen in the Bohr model ground state, \(|\delta_{\pm}| << 1\).   Specifically, at the Bohr radius, \(r_{\text{B}}\),

\begin{equation}
\delta_{\pm} (r = r_{\text{B}}) = \mp \frac{e^4}{2 c^2 \hbar^2} \approx \mp 2.5 \times 10^{-5}.
\label{delta_pm_at_bohr}
\end{equation}

Because the non-Coulomb force due to the spin-orbit interaction is a small perturbation to the Coulomb attractive force, the electron velocity magnitude in the circular orbit is well approximated by its value under Coulomb attraction alone.   Using \(v \approx e/\sqrt{mr}\) in Eq. (\ref{delta_pm_def}) obtains that

\begin{equation}
\delta_{\pm} \approx \mp\frac{1}{2}\frac{v^2}{c^2}.
\label{delta_pm_is_beta_squared}
\end{equation}

\section{Determination of the modified Bohr radii}

The method of Kholmetskii {\em et al.} in Reference 1, for determining the spin-orbit coupling  (and as modified in Reference 3 to include the hidden momentum force), requires determining a modifed Bohr radius that satisfies the Bohr postulate while accounting for the non-Coulomb component of the binding force due to the electron's motion-acquired intrinsic electric dipole moment, \(\mbox{\boldmath$p$}\).  Because the Bohr postulate has been modified here to explicitly include the hidden momentum, it is necessary to derive new modified Bohr radii accordingly, in order to calculate a spin-orbit coupling magnitude where the hidden energy contribution is taken into proper account.  The desired modified Bohr radii for \(n=1\) are solutions of Eq. (\ref{ModBohrPostSimplified}). Solving Eq. (\ref{ModBohrPostSimplified}) for the orbital radius to an accuracy of order \((v/c)^2\) requires that the electron velocity \(v\), a function of the orbital radius, be accurately accounted for to order \((v/c)^2\).  An  expression for the electron velocity under the influence of both the non-Coulomb and Coulomb forces, as needed to solve for \(r\) in Eq. (\ref{ModBohrPostSimplified}), can be found similarly to Eq. (28) of Reference 1, with appropriate modification to account for the halving of the non-Coulomb force due to the presence of hidden momentum.  This results in the removal of the factor of two from the second term on the right hand side Eq. (28) of Reference 1, with result that

\begin{equation}
m v^2 = \frac{e^2}{r}  + \frac{p_{\pm} e}{r^2},  
\label{KetalEq28modified}
\end{equation}

with \(p_{\pm} \equiv \mp |\mbox{\boldmath$p$}|\), where  \(\mbox{\boldmath$p$} = (\mbox{\boldmath$v$} \times \mbox{\boldmath$\mu$})/c\) (Eq. (17) of Reference 1).  With  \(\mbox{\boldmath$\mu$} = -ge\mbox{\boldmath$s$}/2mc\) (with \(g\) the electron g-factor, taken to be 2 herein) directed oppositely to the electron spin vector \(\mbox{\boldmath$s$}\), an outwardly radial  \(\mbox{\boldmath$p$}\) corresponds to antiparallel spin and kinetic orbital angular momenta.

A second equation for \(v^2\) can be obtained by rearranging Eq. (\ref{ModBohrPostSimplified}) and squaring to obtain

\begin{equation}
v^2 = \frac{\hbar^2(1 - \delta_{\pm})^2}{r^2m^2}. 
\label{vsqdfromModBohrPostSimplified}
\end{equation}

Substitution for \(v^2\) in Eq. (\ref{KetalEq28modified}) and straightforward algebra obtains

\begin{equation}
\frac{\hbar^2(1 - \delta_{\pm})^2}{e^2m} = r  + \frac{p_{\pm}}{e}.  
\label{KetalEq28modified3}
\end{equation}

Since \(\delta_{\pm}\) is of order \((v/c)^2\), \((1 - \delta_{\pm})^2 = 1 - 2\delta_{\pm}\) to order \((v/c)^2\).  Using on the left hand side again that \(r_{\text{B}} = \hbar^2/e^2 m\) obtains that, to order \((v/c)^2\),

\begin{equation}
r_{\text{B}}(1 - 2\delta_{\pm}) = r  + \frac{p_{\pm}}{e}. 
\label{r2}
\end{equation}

Defining \(r_{\pm}\) as the orbit radii where the total, including hidden, orbital angular momentum magnitude is \(\hbar\), and accounting for the non-Coulomb component of the binding force to order \((v/c)^2\), and where the spin vector is either parallel (\(r_+\)) or antiparallel (\(r_-\)) to the kinetic orbital angular momentum, then from Eq. (\ref{r2}), to order \((v/c)^2\),

\begin{equation}
r_{\pm} = r_{\text{B}}(1 - 2\delta_{\pm}) -  \frac{p_{\pm}}{e} 
\label{p_pm_1}
\end{equation}

From the Bohr formula and Eq. (\ref{delta_pm_def}) for \(\delta_{\pm}\),

\begin{equation}
r_{\text{B}} \delta_{\pm} = \mp \frac{\hbar^2}{e^2 m} \frac{e^2}{2 m c^2 r}  = \mp \frac{\hbar^2}{2 m^2 c^2 r},
\label{rB_delta_pm}
\end{equation}

and approximating \(\delta_{\pm} (r = r_{\pm})\) by its value at the Bohr radius

\begin{equation}
r_{\text{B}} \delta_{\pm} (r = r_{\pm}) \approx \mp \frac{e^2}{2 m c^2}. 
\label{EmpiricalU}
\end{equation}

To evaluate the third term on the right hand side of Eq. (\ref{p_pm_1}), consider that

\begin{equation}
\frac{p}{e} = \frac{\mu v}{c e} \approx  \frac{1}{c e} \frac{e\hbar}{2 mc} \frac{e}{m^{1/2}r^{1/2}} = \frac{e\hbar}{2 c^2m^{3/2}r^{1/2}},
\label{EmpiricalU}
\end{equation}

so

\begin{equation}
\frac{p_{\pm}(r=r_{\pm})}{e} \approx \frac{p_{\pm}(r=r_{\text{B}})}{e} =  \mp\frac{e\hbar}{2 c^2m^{3/2}} \left(\frac{e^2m}{\hbar^2} \right)^{1/2} = \mp\frac{e^2}{2 m c^2}.
\label{poe_at_rB}
\end{equation}

Equation (\ref{p_pm_1}) may now be written to order \((v/c)^2\) as

\begin{equation}
r_{\pm} = r_{\text{B}}(1 - 2\delta_{\pm}(r=r_{\text{B}})) -  \frac{p_{\pm}(r=r_{\text{B}})}{e} = r_{\text{B}} \pm \frac{3 e^2}{2 m c^2}.
\label{r_pm_final}
\end{equation}

Alternatively, in terms of the fine structure constant,  \( \alpha = e^2/\hbar c \), and electron intrinsic spin magnitude, as \(s = \hbar/2\),

\begin{equation}
r_{\pm} = r_{\text{B}} \pm \frac{3 \alpha^2 \hbar^2}{2 e^2 m} = r_{\text{B}} \left( 1 \pm \frac{s}{\hbar} 3 \alpha^2 \right)
\label{r_pm_alternate}
\end{equation}

\section{Total energy including hidden energy and based on the modified Bohr second postulate}

Let \(U_{\pm}\) represent the total energy of the system of the electron bound in the circular classical orbit around the arbitrarily heavy proton, and with the electron intrinsic spin vector oriented either parallel (for the upper sign) or antiparallel (for the lower sign) to the kinetic orbital angular momentum vector. Then

\begin{equation}
U_{\pm} = K_{\pm} + V_{\pm} + H_{\pm}, 
\label{U_pm_def}
\end{equation}

where \( K_{\pm} \) is the kinetic energy, \(V_{\pm} \) is the potential energy, and \( H_{\pm} \) is the hidden energy.

The kinetic energy of the circular-orbiting electron, under the influence of the combined Coulomb and non-Coulomb forces and including the force due to the time derivative of the hidden momentum, is obtainable from Eq. (\ref{KetalEq28modified}) as

\begin{equation}
K_{\pm} \equiv \frac{m v^2}{2} = \frac{e^2}{2r}  + \frac{p_{\pm} e}{2r^2}.  
\label{KineticGeneral_a}
\end{equation}

With \(p_{\pm}(r=r_{\pm})\) as given by Eq. (\ref{poe_at_rB}), this becomes

\begin{equation}
K_{\pm} = \frac{e^2}{2r_{\pm}}  \mp \frac{e^2}{2{r_{\pm}}^2}\frac{e^2}{2 m c^2},
\label{Kinetic_pm}
\end{equation}

or (using Eq. (\ref{delta_pm_def})),

\begin{equation}
K_{\pm} = \frac{e^2}{2r_{\pm}}\left[1  \mp \frac{1}{r_{\pm}}\frac{e^2}{2 m c^2}\right] = \frac{e^2}{2r_{\pm}}\left[1  \mp \delta_\pm (r_\pm) \right].
\label{Kinetic_pm}
\end{equation}

Since by Eq. (\ref{delta_pm_is_beta_squared}), \(\delta_\pm\) is of order \((v/c)^2\), \(r_{\pm}\) may be approximated as the Bohr radius in the second term on the right hand side of Eq. (\ref{Kinetic_pm}), obtaining that

\begin{equation}
K_{\pm} =  \frac{e^2}{2r_{\pm}}  \mp \frac{e^2}{2}\frac{e^2}{2 m c^2}\left(\frac{e^2 m}{\hbar^2}  \right)^2 =  \frac{e^2}{2r_{\pm}}  \mp \frac{e^8 m}{ 4 c^2 \hbar^4} 
\label{Kinetic_pm_final}
\end{equation}

to order \((v/c)^2\).

The potential energy \(V_{\pm}\) of Eq. (\ref{U_pm_def}) can be expressed as the sum of the Coulomb potential energy and the potential energy of the electron intrinsic magnetic dipole orientation in the magnetic field at the electron in the electron rest frame, due to the proton orbital motion around the electron in its rest frame. As shown in Reference 1, the same expression as Eq. (\ref{V__so_ef1}) here for the orientational potential energy is obtained in the laboratory frame based on the electron's relativistically-acquired electric dipole moment. Therefore, 

\begin{equation}
V_{\pm} \equiv V_{\text{Coulomb}\pm} + V_{s-o \pm},  
\label{V_pm_1}
\end{equation}

with

\begin{equation}
V_{\text{Coulomb}\pm}= -\int_{r=r_{\pm}}^{\infty} \left[\frac{e^2}{r^2} \right] dr =  -\frac{e^2}{r_{\pm}},  
\label{V_a}
\end{equation}

and \(V_{s-o \pm}\) given previously by Eq. (\ref{V_so_final}).  The total potential energy is then

\begin{equation}
V_{\pm} =  -\frac{e^2}{r_{\pm}} \pm \frac{e^8 m}{4c^2\hbar^4}. 
\label{V_pm_final}
\end{equation}

It should be noted that the expression of Eq. (\ref{V_so_final}), as well as that of Eq. (\ref{H_pm}) for the hidden energy, was an approximation based on taking the orbital radius to be exactly the Bohr radius, and the orbital velocity to be that for orbital motion under Coulomb attraction alone.  These approximations cause only order \((v/c)^2\) deviations to energy terms already of order \((v/c)^2\) compared to the kinetic energy and Coulomb potential energy. Therefore the previously-derived values of \(V_{s-o \pm}\) and \(H_{\pm}\), as given by  Eqs. (\ref{V_so_final}) and (\ref{H_pm}) are sufficiently accurate for the present analysis.  

Combining previous results of Eqs. (\ref{H_pm}), (\ref{Kinetic_pm_final}), and (\ref{V_pm_final})  in accordance with the total energy \(U_{\pm}\) definition of Eq. (\ref{U_pm_def}) obtains that 

\begin{equation}
U_{\pm} =  -\frac{e^2}{2r_{\pm}}  \mp \frac{e^8 m}{2 c^2 \hbar^4}. 
\label{Total_pm}
\end{equation}

With \(r_{\pm}\) given by Eq. (\ref{r_pm_final}),

\begin{equation}
-\frac{e^2}{2r_{\pm}}  =  -\frac{e^2}{2}\left[r_{\text{B}} \pm \frac{3e^2}{2m c^2}\right]^{-1} \approx -\frac{e^2}{2r_B}\left[1 \mp \frac{3e^2}{2 m c^2 r_{\text{B}}}\right]= -\frac{e^2}{2r_{\text{B}}} \pm \frac{3e^8m}{4c^2\hbar^4}. 
\label{CoulombE}
\end{equation}

Thus

\begin{equation}
U_{\pm} = -\frac{e^2}{2r_{\text{B}}} \pm \frac{e^8m}{4c^2\hbar^4}, 
\label{Total_pm}
\end{equation}

to order \((v/c)^2\).

The first term on the right hand side of Eq. (\ref{Total_pm}) is simply the energy of the Bohr model in the ground state. The second term on the right hand side is therefore the energy difference due to the spin-orbit coupling, and is seen to be in agreement with Eq. (\ref{V_so_final}) that describes the magnitude of the spin-orbit coupling observed experimentally.  The experimentally-observed spin-orbit coupling has thus been obtained with the inclusion of hidden energy, by modification of the Bohr postulate to account for both kinetic and hidden momentum.  This result can be generalized straightforwardly to other Bohr model energy levels and values of atomic number.

\section{Conclusion}

Hnizdo argues that accounting for a hidden energy that accompanies hidden momentum is necessary to the relativistically covariant description of classical electrodynamic systems. It is found here that the hidden energy of the electron intrinsic magnetic moment in the semiclassical atom is of about the same magnitude as the spin-orbit coupling calculated in the textbook semiclassical analyses that equate the spin-orbit coupling to the potential energy of orientation of the electron intrinsic magnetic moment alone.  If hidden energy is included in the spin-orbit coupling, it appears necessary to modify the Bohr second postulate in a nontrivial fashion if the semiclassical description is to remain in agreement with experiment. This raises a question of whether the de Broglie wavelength expression should be so modified as well.  That is, with the de Broglie wavelength given as \(\lambda = h/p\), where \(h\) is the Planck constant and \(p\) is the momentum magnitude, and noting that the Bohr radii are ones with orbit circumference of whole multiples of the de Broglie wavelength, should \(p\) consist properly of hidden as well as kinetic momentum?  It seems that perhaps it should, as there is already support in the literature for the influence of hidden momentum in the Dirac equation \cite{Shockley1968}.

\section*{Acknowledgment}

I thank Dr. Kholmetskii, Dr. Missevitch, and Dr. Yarman for interesting discussions on this topic.  It is Dr. Kholmetskii's observation that explicitly including the hidden momentum in the Bohr second postulate implies a modification of the de Broglie wavelength definition.

\bibliographystyle{plain}


\end{document}